\documentclass[letterpaper,10pt,conference]{ieeeconf}
\usepackage{graphics}
\usepackage{graphicx}
\usepackage{tabularx}
\usepackage{multirow}
\usepackage{float}
\usepackage{cite}
\usepackage{subfig}
\usepackage{amsmath,amssymb,amsfonts}
\usepackage[dvipsnames]{xcolor}
\usepackage{array,booktabs}
\usepackage{bm}
\usepackage{cite}
\usepackage{scalerel}

\IEEEoverridecommandlockouts
\begin{document}

\title{\LARGE \bf On the use of generative deep neural networks to synthesize artificial multichannel EEG signals}

\author{Ozan \"{O}zdenizci$^{\star,\dagger}$ and Deniz Erdo\u{g}mu\c{s}$^{\star}$%
\thanks{$^{\star}$Cognitive Systems Laboratory, Department of Electrical and Computer Engineering, Northeastern University, Boston, MA, USA.}%
\thanks{$^{\dagger}$Institute of Theoretical Computer Science, Graz University of Technology, Graz, Austria. E-mail: ozan.ozdenizci@igi.tugraz.at}%
\thanks{This work is partially supported by NSF (IIS-1149570, CNS-1544895, IIS-1715858), DHHS (90RE5017-02-01), and NIH (R01DC009834).}%
}

\maketitle
\thispagestyle{empty}
\pagestyle{empty}


\begin{abstract}Recent promises of generative deep learning lately brought interest to its potential uses in neural engineering. In this paper we firstly review recently emerging studies on generating artificial electroencephalography (EEG) signals with deep neural networks. Subsequently, we present our feasibility experiments on generating condition-specific multichannel EEG signals using conditional variational autoencoders. By manipulating real resting-state EEG epochs, we present an approach to synthetically generate time-series multichannel signals that show spectro-temporal EEG patterns which are expected to be observed during distinct motor imagery conditions.\end{abstract}

\section{Introduction}

Electroencephalography (EEG) based brain-computer interface (BCI) technologies provide a direct neural communication channel for people with neuromuscular disabilities by conveying user intent decoded from non-invasive brain activity, to an external controllable computer/machine interface \cite{Schalk:2004}. A common research problem that remains of important interest in developing EEG-based BCI systems is to tackle with inter- and intra-recording variabilities in EEG across- or within-users. Such phenomena caused by the non-stationary nature of EEG signals influence the robustness of EEG intent decoding pipelines that are learned with limited amount of training data. In this context, previous studies put forward the idea of generating artificial EEG epochs for data augmentation using the few available training trials \cite{Lotte:2015}.

Recent progress in generative deep neural networks drew significant attention in this regard. Particularly in image processing, deep generative models demonstrated their capabilities to produce highly realistic artificial images \cite{Kingma:2013,Goodfellow:2014}. One preliminary architecture is \textit{variational autoencoders (VAE)} \cite{Kingma:2013}, where conventional feature autoencoders were extended into learning a pair of a stochastic encoder and a deterministic decoder network that re-generates images simultaneously, yielding the capability to draw novel samples through the probabilistic nature within the model. Later with \textit{generative adversarial networks} (GAN), a generator model is learned to synthesize data samples using latent noise sampled from a noise distribution, while a discriminator model is simultaneously trained to distinguish real data samples from artificially generated ones \cite{Goodfellow:2014}. As an outcome of this competing training approach, generative models that can produce realistic data samples can be learned.

Consistently with these advancements, recent EEG deep learning applications began to explore whether realistic EEG signals can also be artificially generated. In this paper, we first review recent advances on artificial EEG signal augmentation using generative deep learning. We later present our experimental explorations on synthetically generating condition-specific EEG signals with conditional VAEs. In our data analyses, from a different perspective, we aim to manipulate real resting-state EEG epochs into synthetic multichannel EEG time-series that can demonstrate condition-specific spectro-temporal characteristics of motor imagery.

\section{Related Work: Generative EEG Deep Learning}

Deep neural networks have been widely explored as generic feature learning machines tailored to specific EEG classification tasks. Most studies use convolutional model architectures which are shown capable of capturing spatial, temporal and spectral dynamics from multichannel raw input EEG signals for classification \cite{Schirrmeister:2017,Lawhern:2018,Ozdenizci:2020}. From the recent generative EEG deep learning aspect, majority of studies explored GAN and deep convolutional GAN (DCGAN) models as compared to VAE based approaches. Capabilities of GANs to learn unlabeled EEG dynamics were broadly studied in \cite{Fahimi:2019}. By presenting time-series and topographical visualizations of generated signals, successful model learning processes were illustrated. Another application of GANs studied generating synthetic EEG epochs that demonstrated time-series characteristics of EEG recorded during epileptic seizures \cite{Pascual:2019}. Differently, \cite{Zhang:2018} used conditional DCGANs with network input architectures of Wavelet time-frequency transform images of EEG to capture event-related de-/synchronization (ERD/ERS) \cite{Pfurtscheller:1997} patterns from these images. Learning such an artificial time-frequency representation image generator yielded beneficial results in motor imagery task classification. Similarly \cite{Fahimi:2020} and \cite{Roy:2020} utilized GANs towards generating artificial motor imagery EEG signals by using multiple class-specific trained models. \cite{Fahimi:2020} emphasized on boosting motor imagery classification results with data augmentation, whereas \cite{Roy:2020} reported imagery relevant beta-power time-frequency variations captured by the generative networks. \cite{Aznan:2019} explored DCGANs and VAEs on steady-state visual evoked potentials and demonstrated learned power spectral density patterns. Another study \cite{Abdelfattah:2018} which comparatively explored VAEs also favored a recurrent GAN model based on evaluations via task decoding results with artificially augmented training data. From a different perspective, \cite{Corley:2018} used GANs to artificially up-sample spatial channel resolution of EEG for recovering recordings with less sensors.

Another ongoing line of work explored Wasserstein GAN (WGAN) models in EEG data generation. One of the earlier examples \cite{Hartmann:2018} focused on single-channel EEG. The study distinctively explored temporal and spectral characteristics of artificial EEG epochs by also showing spectral power differences between two classes of generated data (i.e., motor execution versus rest). Similarly \cite{Luo:2020a} explored WGAN models in comparison to VAEs for emotion-elicited artificial EEG signal generation, where data augmentation is performed in an extracted power spectral density (PSD) feature space and evaluated by beneficial emotion recognition classification performances while temporal frequency dynamics of generated signals were not explored. A variation of conditional WGANs were also used for rapid serial visual presentation EEG responses \cite{Panwar:2019}, yet models were again only evaluated based on classification performance.

\section{Methods}

\subsection{Conditional Variational Autoencoders (cVAE)}

Consider an EEG data set $\{(\bm{X}_i,y_i)\}_{i=1}^{n}$ with $\bm{X}_i$ denoting the time-series EEG data matrix at trial $i$, $y_i\in\{1,\ldots,L\}$ indicating the class label, and $\bm{c}_i$ denoting the $L$-dimensional one-hot encoded vector version of $y_i$. VAEs \cite{Kingma:2013} construct generative models based on a stochastic encoder and a deterministic decoder network pair. The encoder estimates a variational posterior distribution $q_{\phi}(\bm{z}\vert\bm{X}) \sim \mathcal{N}(\bm{\mu_z},\bm{\sigma_z})$ for a latent representation $\bm{z}$, by learning the mean and covariance parameter vectors $\bm{\mu_z}$ and $\bm{\sigma_z}$. Using these distribution parameter estimates, novel $\bm{z}$ observations can be sampled. This sampled $\bm{z}$ is used by the decoder to reconstruct $\bm{\hat{X}}$.

In cVAEs \cite{Sohn:2015}, the decoder is further conditioned on $\bm{c}$ as an additional input. Here the encoder is hypothesized to learn a $\bm{c}$-invariant variational posterior, and during inference this further introduces the flexibility to manipulate $\bm{c}$ and generate condition-specific novel samples using the learned distribution. Objective of the model is to approximate a latent space where samples $\bm{z}$ are more likely to reconstruct $\bm{X}$. This corresponds to training the model based on the well known \textit{evidence lower bound (ELBO)} \cite{Kingma:2013,Sohn:2015}. The loss function to be minimized (negated ELBO) is given by:
\begin{equation}
\mathcal{L}(\bm{X}) = \mathbb{E}\bigl[-\log p_{\theta}(\bm{X} \vert \bm{z},\bm{c}) \bigr] + KL\bigl( q_{\phi}(\bm{z}\vert\bm{X})\vert\vert p(\bm{z}) \bigr).
\label{eq:cvae}
\end{equation}

Here the first term corresponds to expected log-likelihood maximization based on the reconstructions (i.e., minimizing mean-squared error for $\bm{\hat{X}}$). The second term represents minimizing the Kullback-Leibler (KL) divergence as a distance measure between two distributions, which are the variational posterior and the true latent distribution. Conventionally $p(\bm{z})$ is assumed to have a standard normal distribution.

\renewcommand{\arraystretch}{1.1}
\begin{table}[t!]
	\caption{cVAE model architecture used for EEG synthesis (BN: batch norm, ELU: exponential linear unit).}
	\label{tab:cvae}
	\begin{tabular}{l l l l l}
		\hline
		& \textbf{Layer} & \textbf{Kernel Size} & \textbf{Output Dim.} \\
		\hline
		\parbox[t]{.5mm}{\multirow{5}{*}{\rotatebox[origin=c]{90}{\textbf{Encoder}}}} & Input EEG: $\bm{X}$ &  & $(1,15,400)$ \\
		& Conv2D + BN + ELU & $5 \times (1,40)$ & $(5,15,400)$\\
		& Conv2D + BN + ELU & $5 \times (15,1)$ & $(5,1,400)$ \\
		& Mean Pooling + Flatten & $(1,2)$ & $(1,1000)$ \\
		& Dense & $2 \times (1000,10)$ & $2\times(1,10)$ \\
		\hline
		& Sample $\bm{z}\sim\mathcal{N}(\bm{\mu_z},\bm{\sigma_z})$ & & $(1,10)$ & \\
		& Concatenate $\bm{c}$ to $\bm{z}$ & & $(1,13)$ & \\
		\hline
		\parbox[t]{.5mm}{\multirow{5}{*}{\rotatebox[origin=c]{90}{\textbf{Decoder}}}} & Dense & $(13,1000)$ & $(1,1000)$ \\
		& Reshape + Upsampling & $(1,2)$ & $(5,1,400)$ \\
		& Deconv2D + BN + ELU & $5 \times (15,1)$ & $(5,15,400)$ \\
		& Deconv2D + BN + ELU & $5 \times (1,40)$ & $(1,15,400)$ \\
		& Output EEG: $\bm{\hat{X}}$ & & $(1,15,400)$ \\
	    \hline
	\end{tabular}\vspace{-0.2cm}
\end{table}

\subsection{Experimental Data and Model Learning}

We used the PhysioBank EEG Motor Movement/Imagery Dataset \cite{Goldberger:2000}, originally consisting of single session offline EEG recordings from 109 subjects while 64-channel EEG were recorded at 160 Hz sampling rate. Subjects were performing visual cue based \cite{Schalk:2004} motor execution or motor imagery tasks of the right-, left-hand, both fists or both feet in randomized order. Each trial lasted four seconds with inter-trial resting periods of same duration. At the very beginning of recording sessions, subjects' one-minute eyes-open and eyes-closed resting-state EEG were also recorded. We only considered 100 subjects' data with proper trial timestamp alignments of recordings, and the trials with right, left hand and both feet motor imagery tasks \cite{Pfurtscheller:2006}. Henceforth we denote these three motor imagery class conditions by RIGHT, LEFT and FEET. For cVAE model learning this yielded a total of 6300 trials (i.e., 21 trials per class and subject). To investigate class-specific spectro-temporal (de-)modulation dynamics of EEG, we epoched each trial from the 0.5 sec pre-cue to 2 sec post-cue time interval with the cue indicating the motor imagery onset. Trials were epoched based on the cues since there were no available electromyography recordings that annotate any actual muscle activation onsets. We only used the 15 EEG channels over the sensorimotor strip forming a rectangular region around electrode Cz (Fig~\ref{fig:originalR}). This resulted in cVAE inputs with 15 channels by 400 time samples. For pre-processing, EEG data were common average referenced and 4-30 Hz bandpass filtered with a causal third order Butterworth filter.

Implementations were done in Tensorflow with the Keras API. We train a cVAE using a pooled training set of 5400 trials, and a validation set with the remaining 900 including trials from all subjects and also stratified by class labels. Parameter updates were performed once per batch with Adam optimizer using 50 training trials per batch, for at most 100 epochs with validation set loss based early stopping. Using a parameter grid search towards minimum validation set loss, we determined $\bm{z}$ to be 10-dimensional and used 5 kernels per convolution layer. Kernel lengths were chosen in consistency with the signal sampling rate and frequency range \cite{Lawhern:2018} (see Table~\ref{tab:cvae} for specifications with 34,214 total parameters to be learned). Condition vector $\bm{c}$ provided to the decoder was three dimensional, formed as a one-hot encoded vector of $y$ (e.g., RIGHT: $\bm{c}=[1,0,0]$).

\begin{figure}[t!]
\includegraphics[width=0.49\textwidth]{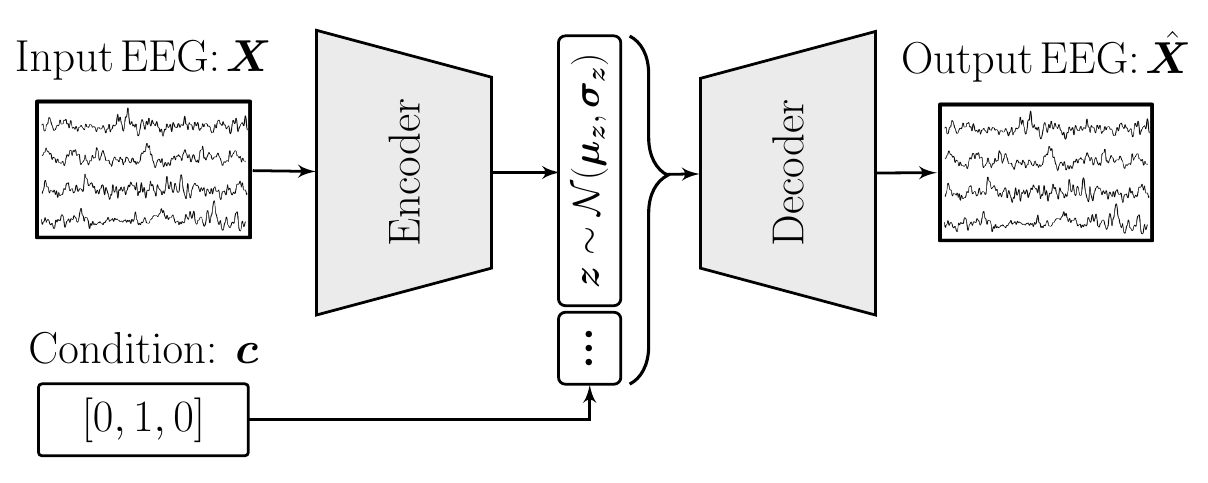}
\caption{Class-specific synthetic data generation with cVAEs. At model training time, input EEG are provided along with their associated class condition $\bm{c}$. For data generation, resting-state EEG are provided along with a preferred $\bm{c}$.}
\label{fig:cvaegeneration}\vspace{-0.3cm}
\end{figure}

\subsection{Manipulating Resting-State EEG Epochs}

After training the cVAE, we aimed to generate condition-specific artificial EEG trials by manipulating real resting-state EEG epochs. Using each subject's one-minute eyes-open resting recordings, we extracted EEG epochs of 2.5 sec duration such that the resting EEG segments matched the cVAE input dimensionality. We discarded the first and last five seconds of these one-minute resting recordings, and split the remaining 50 seconds of data into 2.5 sec long non-overlapping segments. This resulted in 20 resting-state EEG epochs per subject (i.e., 2000 epochs in total).

Each resting-state epoch was then used as an input for the trained cVAE model, yielding a newly sampled latent $\bm{z}$ to be fed into the decoder. Since resting EEG epochs did not have a ground-truth condition $\bm{c}$ with respect to the training data set, we could simply use these epochs to provide a user-specific EEG template to generate new data. Specifically we manipulated $\bm{c}$, which will be provided to the decoder alongside $\bm{z}$, in a way that it generated an $\bm{\hat{X}}$ with the desired condition-specific EEG dynamics. An illustration is provided in Fig~\ref{fig:cvaegeneration}, where a condition vector $\bm{c}=[0,1,0]$ is chosen to generate an EEG epoch which would belong to class LEFT. 

\section{Experimental Results and Discussion}
\label{sec:experiments}

Artificial EEG epochs were generated in the form of multichannel time-series signals. To investigate their spectro-temporal dynamics, we calculated time-frequency representations (TFRs) for all the epochs per channel and trial. TFRs were calculated in 4-30 Hz frequencies and [-0.5,2] sec pre/post-cue time interval, using multi-taper spectral analysis \cite{Percival:1993} with two Slepian tapers, a 500 ms time window and 50 ms window step size. Each 1 Hz frequency bin were then baseline-referenced to indicate the signed percentage power change (ERD/ERS) \cite{Pfurtscheller:1997} with respect to the average power within the [-0.5,0] sec pre-cue baseline. This resulted in ERD/ERS maps for each electrode and trial.

\begin{figure}[t!]
\includegraphics[clip,trim=0 0 0 0.3cm, width=0.49\textwidth]{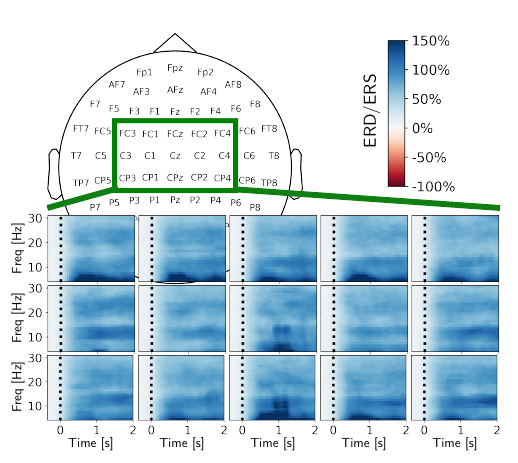}
\caption{Our 15-channel EEG layout and TFRs per electrode showing the averaged ERD/ERS maps of real resting-state EEG. Color bar denotes \% power change from the baseline.}
\label{fig:originalR}\vspace{-0.3cm}
\end{figure}

\begin{figure}[!t]
\centering
\subfloat[Artificial motor imagery epochs for condition RIGHT]{\includegraphics[clip,trim=0 0 0.1cm 0.2cm, width=0.49\textwidth]{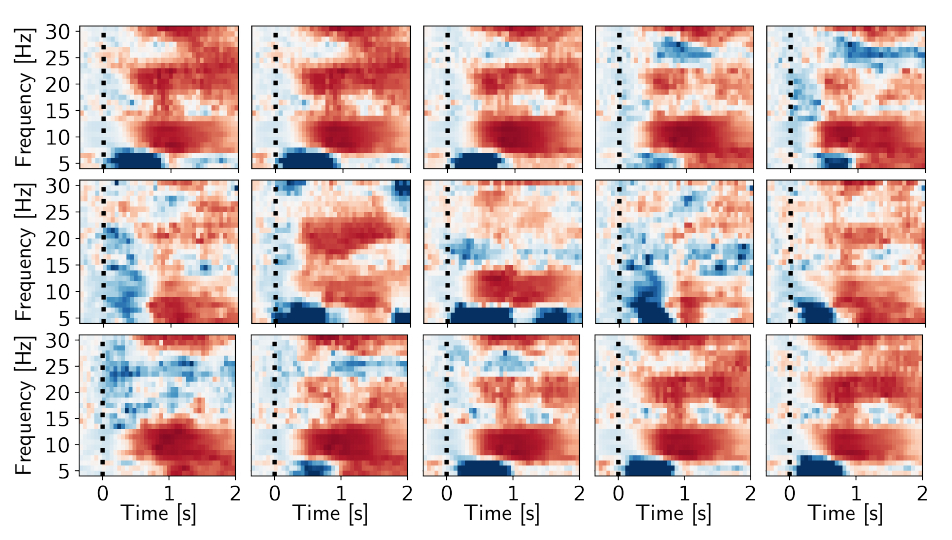}\vspace{-0.15cm}\label{fig:generatedR}}\\\vspace{-0.15cm}
\subfloat[Artificial motor imagery epochs for condition LEFT]{\includegraphics[clip,trim=0 0 0.1cm 0.2cm, width=0.49\textwidth]{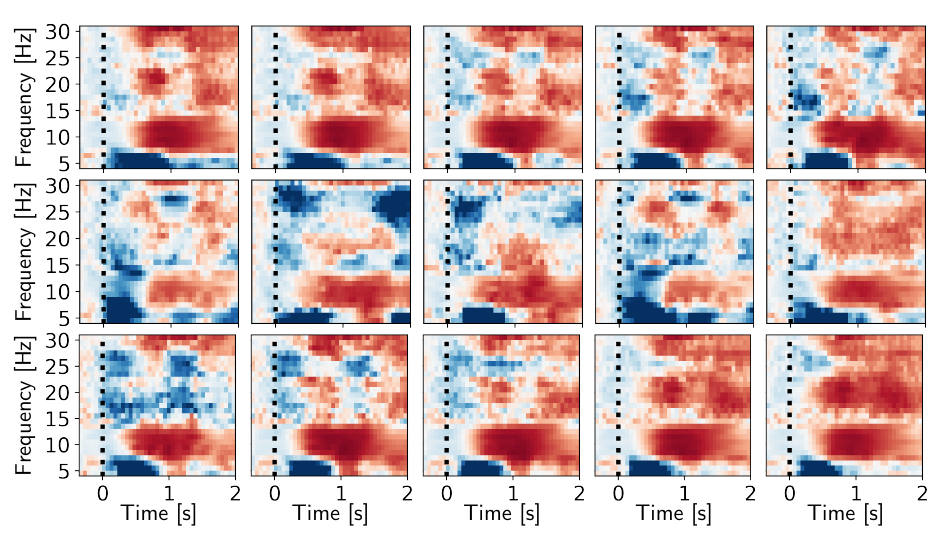}\vspace{-0.15cm}\label{fig:generatedL}}\\\vspace{-0.15cm}
\subfloat[Artificial motor imagery epochs for condition FEET]{\includegraphics[clip,trim=0 0 0.1cm 0.2cm, width=0.49\textwidth]{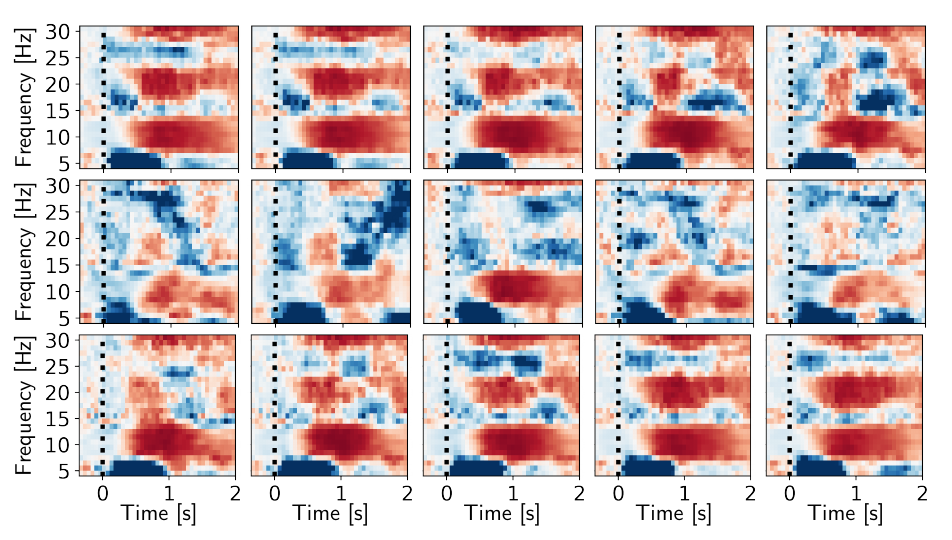}\label{fig:generatedF}}%
\caption{ERD/ERS patterns averaged across 2000 synthetic class-specific epochs. Color bar and the 15-channel layouts are as depicted in Fig.~\ref{fig:originalR}. Dotted lines indicate the cue onset.}
\vspace{-7pt}
\label{fig:artificialgen}
\end{figure}

Fig~\ref{fig:originalR} depicts the layout of 15 EEG channels, as well as the ERD/ERS maps of real resting EEG epochs averaged across 2000 maps from all subjects. Resting-state epochs indicate no distinct patterns across frequencies as expected. We believe that the overall slight power increase can be due to an edge artifact from the baseline period. Fig~\ref{fig:artificialgen} illustrates the ERD/ERS maps of artificial EEG epochs for the three conditions, where each map presents averaged TFRs across 2000 maps obtained by condition-specific manipulation of resting data from all subjects. We primarily observe that neutral resting EEG recordings can be transformed into time-series EEG signal epochs which demonstrate the $\alpha$- (8-15 Hz) and $\beta$- (20-30 Hz) frequency band ERD phenomena spectrally \cite{Pfurtscheller:1997,Pfurtscheller:2006} following the cue onset.

In Fig~\ref{fig:artificialgen}, while we do not observe strong class-specific (de-)modulation differences at the five prefrontal (FP) and five centro-parietal (CP) electrodes (i.e., at the top and bottom rows), instead we observe strong class-specific differences across the central strip of electrodes. Particularly, for class condition RIGHT (Fig~\ref{fig:generatedR}) we observed a stronger $\beta$-band ERD pattern at the electrodes placed on the left hemisphere, for class LEFT (Fig~\ref{fig:generatedL}) a cleaner $\alpha$-band ERD at the right hemisphere electrode C4 \cite{Pfurtscheller:1997}, and for FEET (Fig~\ref{fig:generatedF}) a strong $\alpha$-band ERD presence at the central electrode Cz which is in consistency with the distinct central patterns usually observed during feet movement imagery \cite{Pfurtscheller:2006}.

\begin{figure}[t!]
\centering
\subfloat[\% change in $\alpha$-power]{\includegraphics[clip, trim=0 0 0.5cm 0.5cm, width=0.24\textwidth]{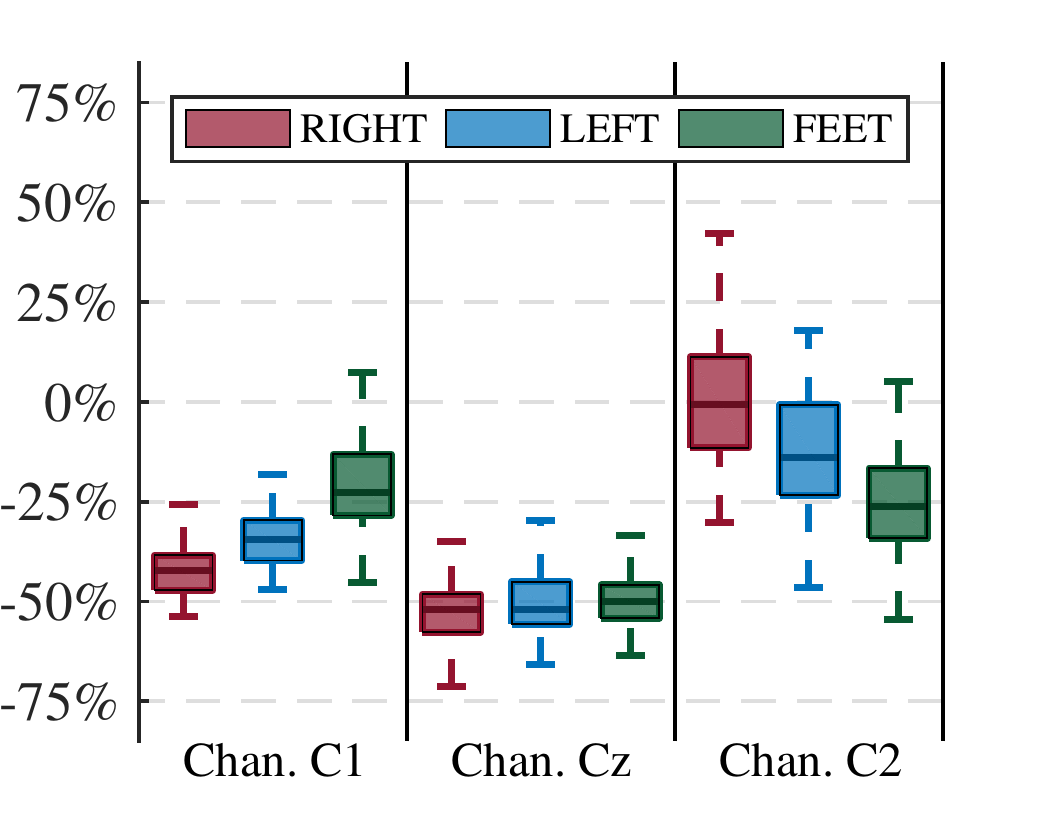}\label{fig:alphachange}}
\subfloat[\% change in $\beta$-power]{\includegraphics[clip, trim=0 0 0.5cm 0.5cm, width=0.24\textwidth]{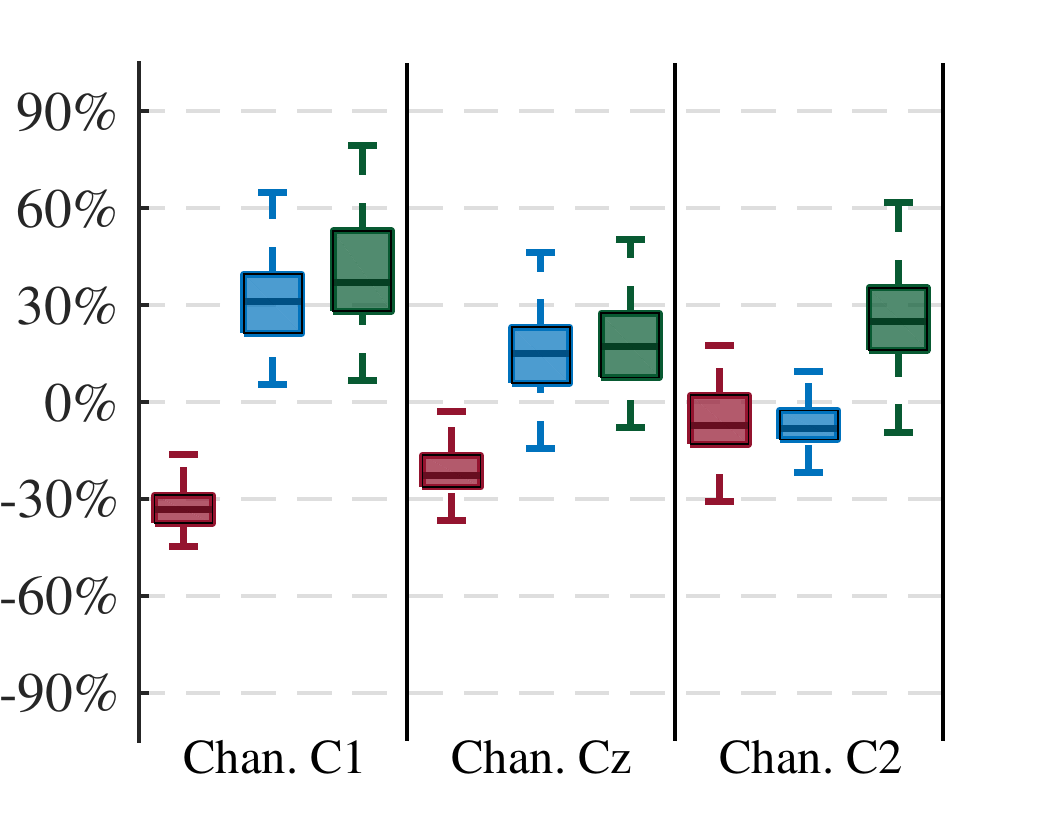}\label{fig:betachange}}
\caption{$\alpha$- and $\beta$-band ERD/ERS distributions of artificial EEG epochs across subjects. Each box represents a distribution of power change values across 100 subjects (i.e., center marks show the median, upper/lower bounds show first/third quartiles, and dashed lines show extreme values).}
\label{fig:bpchanges}
\end{figure}

Lastly, Fig~\ref{fig:bpchanges} depicts the artificially generated variability of bandpower changes across subjects. Considering our use of subject-specific resting EEG templates, we obtain and demonstrate subject- and class-specific averages of percentage $\alpha$- and $\beta$-power changes. For each subject, an average percentage bandpower change value was calculated over the [0.5,1.5] post-cue time interval and the particular frequency band of the TFRs, and over 20 subject- and class-specific artificial EEG epochs. Fig~\ref{fig:bpchanges} reveals consistent class-relevant differences. We observe that the distribution of $\alpha$-ERD is stronger on the right hemisphere (C2) for condition LEFT than RIGHT and vice versa for channel C1, whereas the variability of $\alpha$-ERD at channel Cz does not differ between classes. In Fig~\ref{fig:betachange} we observe a strong contralateral $\beta$-ERD distribution for RIGHT, as anticipated from Fig~\ref{fig:generatedR}.

\section{Conclusion}

We present condition-specific spectro-temporal dynamics of artificial motor imagery EEG epochs synthesized through a cVAE. We believe that these generated data still needs to be (1) validated with comparisons to real motor imagery recordings on a subject-specific basis, and (2) compared with other deep data synthesis models \cite{Ozdenizci:2019,Fahimi:2020,Roy:2020}. Conditional models introduce a semi-supervised flexibility to consider any arbitrary EEG epoch as a user-specific template. This can be a long-term motivation for BCIs or diagnostic models \cite{Pascual:2019}, where generative models can provide a systematic way to manipulate recorded EEG for finetuning decoding models.

\end{document}